\documentclass[%
 aip,
 amsmath,amssymb,
 reprint,%
floatfix]{revtex4-1}
\usepackage{lipsum}
\usepackage{appendix}
\usepackage{color}
\usepackage{soul}
\usepackage[normalem]{ulem}
\usepackage{graphicx}
\usepackage[nice]{nicefrac}

\begin{document}
\title{Capture of rod-like molecules by a nanopore: \\ defining an "orientational capture radius"}
\author{Le Qiao}
\email{lqiao@uottawa.ca}
\author{Gary W. Slater}
 \email{gslater@uottawa.ca}
\affiliation{ Department of Physics, University of Ottawa, Ottawa, Ontario, Canada, K1N 6N5}

\date{\today}
\begin{abstract}
Both the translational diffusion coefficient $D$ and the electrophoretic mobility $\mu$ of a short rod-like molecule (such as dsDNA) that is being pulled towards a nanopore by an electric field should depend on its orientation. Since a charged rod-like molecule tends to orient in the presence of an inhomogeneous electric field, $D$ and $\mu$ will change as the molecule approaches the nanopore, and this will impact the capture process. We present a simplified study of this problem using theoretical arguments and Langevin Dynamics simulations. In particular, we introduce a new \textit{orientational capture radius} which we compare to the capture radius for the equivalent point-like particle, and we discuss the different physical regimes of orientation during capture and the impact of initial orientations on the capture time.
\end{abstract}

\maketitle

\section{Introduction}
\label{sec:Introduction}

Field-driven translocation through a nanopore can be used to analyze biomolecules like microRNA, DNA and proteins\cite{beamishIdentifyingStructureShort2017,wanunuRapidElectronicDetection2010,kowalczykDetectionLocalProtein2010,wadugeNanoporeBasedMeasurementsProtein2017,beamishProgrammableDNANanoswitch2019,charronPreciseDNAConcentration2019,heFastCaptureMultiplexed2019,bandarkarHowNanoporeTranslocation2020}, and new methodologies are continuously being  proposed  to  enhance  the  performance of the related devices, \textit{e.g.} improving the capture by pre-confining the DNA with a nanoporous filter\cite{lamEntropicTrappingDNA2019,briggsDNATranslocationsNanopores2018}, controlling the translocation time by coating the nanopore with a lipid bilayer\cite{yuskoControllingProteinTranslocation2011,eggenbergerFluidSurfaceCoatings2019}, achieving multiplexed detection using DNA-based labels or carriers\cite{beamishIdentifyingStructureShort2017,plesaVelocityDNATranslocation2015,charronPreciseDNAConcentration2019} \textit{etc.} Unlike spherical objects, highly charged dsDNA molecules can either deform and stretch (if their contour length $L$ is much larger than their persistence length $L_p$) or simply orient (if $L<L_p$) during the capture process\cite{farahpourChainDeformationTranslocation2013,vollmerTranslocationNonequilibriumProcess2016,sarabadaniIsofluxTensionPropagation2014,buyukdagliTheoreticalModelingPolymer2019}; for dsDNA, $L_p \sim 50\,nm$ or $\sim 150\,bp$ under typical salt conditions\cite{rosensteinIntegratedNanoporeSensing2012}. The impact of the coupling between the dsDNA conformation/orientation and its dynamics is often neglected.  Another example is the translocation of a rod-shaped virus \cite{mcmullenStiffFilamentousVirus2014,ventaGoldNanorodTranslocations2014}:
the tobacco mosaic virus (TMV)\cite{alonsoPhysicsTobaccoMosaic2013,kochNovelRolesWellknown2016,schmatullaMicromechanicalPropertiesTobacco2007} is a charged rigid rod of length $\sim 300\,nm$ and diameter $\sim 15\,nm$ with a persistence length $>10$ times its length. 

Previous studies of the translocation of charged rod-like objects mainly focused on how rods enter a nanopore and then translocate\cite{mcmullenStiffFilamentousVirus2014,wuTranslocationRigidRodShaped2016,d.y.bandaraCharacterizationFlagellarFilaments2019,buyukdagliTheoreticalModelingPolymer2019}. In this short paper, we examine the capture of a short rod-like object by the field gradient extending outside a nanopore, and in particular the impact of rod orientation on the capture process. Describing the orientation of the rod using a local order parameter, we conclude that the physics of the problem is related to a new length scale $R_\theta$ that characterizes the radial position where orientation starts. Since $R_\theta$ is smaller than the standard capture radius, rods drift faster than they can orient, with potential impact on capture rates.

\section{Theoretical analysis}
\label{sec:Theory_models}

\begin{figure}[ht]
\centering
\includegraphics{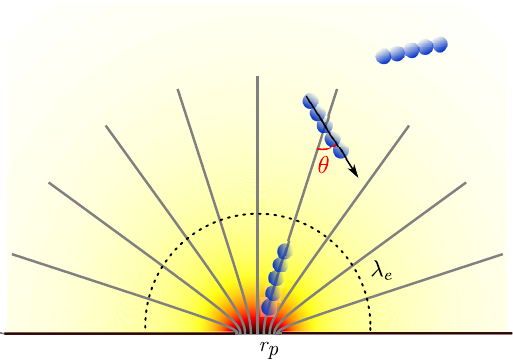}
\caption{A schematic view of the nanopore system. The background colours code for the electric field strength (red means higher fields). The dashed line depicts the capture radius $\lambda_e$ while the grey lines depict field lines.The angle between the rod and the electric field line is $\theta$}
\label{Fig_system}
\end{figure}

As discussed in the next subsection, the field lines are radial and the field intensity decays as $E  \sim 1/r^2$ at large distances from a nanopore -- see Fig.~\ref{Fig_system}. A rod-like molecule thus feels a torque and orients in the resulting field gradient to minimize its local energy. We will characterize the mean orientation using the order parameter
\begin{equation}
\label{orderpara}
    \Theta = \tfrac{1}{2} \left[ {3\langle {\cos^2 \theta}\rangle-1} \right],
\end{equation}
where $\theta$ is the angle between the molecule's principal axis and the local field direction. Note that $\Theta=1$ for perfect alignment, while $\Theta=0$ for random orientation ($r \rightarrow\infty$).

\subsection{The electric field and forces}
\label{sec:field}

For radial distances $r$ much larger than the pore radius $r_p$, the electric field outside the pore can be represented by the point-charge approximation\cite{qiaoVoltageDrivenTranslocationDefining2019,wanunuElectrostaticFocusingUnlabelled2010}
\begin{equation}
    \label{eq:EInfinity}
   {E}(r) \approx -{r_e \Delta V}/{r^2} ,
\end{equation} 
where $\Delta V$ is the potential difference across the system and the length scale  $r_e=r_p/(\frac{2l}{r_p}+\pi)$  describes the pore's size and aspect ratio ($l$ is the pore length). This approximation will be used in our theoretical analysis, while an analytical solution to Laplace's equation\cite{farahpourChainDeformationTranslocation2013,kowalczykModelingConductanceDNA2011}  will be used in the simulations. 

As described in\cite{grosbergDNACaptureNanopore2010}, $Q=k_BT \mu /D$ is the effective DNA electrophoretic charge, where $\mu$ and $D$ are the mobility and diffusion coefficient of the DNA in free solution. The standard definition of the capture radius is the length scale $r=\lambda_e$ where the analyte's potential energy $QV(r)=k_BT$; this gives
\begin{equation}
\label{eq:lambda_definition}
    \lambda_e =\frac{Q \Delta V }{k_BT}~ r_e.
\end{equation}
Note that we will use $\lambda_e$ to measure the amplitude of the applied electric forces; for instance, the velocity of the rod can then be written simply as $v(r)=\lambda_e D/r^2$.

For the simulations, we chose the following dimensionless parameters: a rod of length $L=\tfrac{10}{3}\,r_p$, a pore aspect ratio $l/r_p=2$ (giving $r_e=\tfrac{r_p}{4+\pi}$) and fields in the range of $\lambda_e=200-1000 \,r_p$. As a guide, if one were to map this simulation onto the dynamics of a short $L=100\,bp$ (or $\approx 34\,nm$ long) dsDNA, the pore size would be $r_p \approx {10~nm}$, the effective rod charge would be $Q=k_BT \mu/D \approx 70\,e$, and the voltages would be in the range $\Delta V \approx 0.5 \! - \! 2.5\,V$.

\subsection{Static orientation in the field gradient}
\label{sec:static_theory}
\begin{figure}[ht]
\centering
\includegraphics{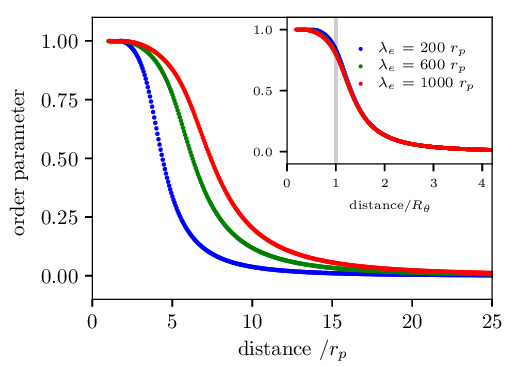}
\caption{Static order parameter $\Theta(r)$ \textit{vs} scaled distance $r/r_p$ for a rod of length $L=\tfrac{10}{3}\,r_p$ and different field intensities $\lambda_e$ (in units of $r_p$), as obtained from numerical integration of eq. \ref{eq_avarageCos}. Inset: Same data with the $x$-axis now rescaled using $R_\theta$, with $R_\theta(\lambda_e=200\,r_p)=3.3\,r_p$, $R_\theta(600\,r_p)=4.8\,r_p$ and $R_\theta(1000\,r_p)=5.7\,r_p$. }
\label{Fig_static_rotation}
\end{figure}

We consider a uniformly charged rigid rod of length $L$ whose centre of mass (CM) is at position $r$, and we assume that it is in orientational equilibrium in the potential $V(r)$. Its potential energy in this radial field is
\begin{equation}
\label{eq_rotation_V_Gary2}
\frac{\Psi_{\theta}(r)}{k_BT}=\frac{\lambda_e}{L}\cdot \int^{+L/2}_{-L/2}~\frac{\mathrm{d}z}{\sqrt{r^2+z^2+2rz\cos{\theta}}} , 
\end{equation}
where $z$ is the distance between a charge along the rod and the centre-of-mass of the rod. Note that the potential energy of the rod depends only on the distance $r$ and the angle $\theta$ because the field in eq.~2 is radial. The orientational potential energy for the rod is thus
\begin{equation}
\label{eq_rotation_V}
\delta \Psi_\theta(r)=\Psi_{\theta=0}(r)-\Psi_{\theta}(r),
\end{equation}
and the corresponding mean orientation is given by\cite{slaterNewBiasedReptationModel1985,slaterConstructionApproximateEntropic1994}
\begin{equation}
\label{eq_avarageCos}
  \langle \cos^2 \theta (r) \rangle= \frac{\int^{\pi}_{0} \cos^2 \theta \sin \theta \exp{\left(-\delta \Psi_\theta(r)/k_BT\right)} \mathrm{d}\theta}{\int^{\pi}_{0} \sin \theta \exp{\left(-\delta \Psi_\theta(r)/k_BT\right)}\mathrm{d}\theta}.
\end{equation}
Although these integrals cannot be done in closed   form, they can be computed numerically to obtain the order parameter $\Theta(r)$ for different values of the nominal capture radius $\lambda_e$ -- see Fig.~\ref{Fig_static_rotation}. As expected, $\Theta(r)$ drops quickly with distance because the field gradient decays as $r^{-3}$. We notice that distances much smaller than $\lambda_e$ are needed to obtain substantial orientation: in other words, the DNA rod is "captured" much before it orients. 

When $r \! \gg \! L$, the asymptotic form of eq.~\ref{eq_avarageCos} is 
\begin{equation}
    \Theta(r) \approx {\lambda_eL^2}/{60~r^3} \equiv \left({R_\theta}/{r}\right)^3,
\end{equation}
where the length scale $R_\theta$ will be called the \textit{orientational capture radius}. Since $\lambda_e \sim Q \sim \mu/D$, it scales like
\begin{equation}
\label{eq:orientation_radius}
    R_\theta = \left(\tfrac{1}{60}\lambda_eL^2\right)^{1/3}\sim \left(L^2 \Delta V / D\right)^{1/3}.
\end{equation}
The inset of Fig.~\ref{Fig_static_rotation} shows the same data, but with $r$ now rescaled using $R_\theta$. The curves collapse, except (weakly) at very short distances. The orientational capture radius $R_\theta=\sqrt[3]{\lambda_e L^2 /60}$~ is thus the length scale describing the decay of the order parameter. Importantly, we have $R_\theta \ll \lambda_e$ since $L \ll \lambda_e$. Given that $D \sim 1/L$ for a rod, this relation also predicts that $R_\theta \propto L \Delta V^{\nicefrac{1}{3}}$, where the $\nicefrac{1}{3}$-scaling comes directly from the field gradient. Note that we chose three dimensionless field intensities $\lambda_e > 60L$ to insure that $R_\theta \ge L$.

\subsection{Scaling analysis}
\label{sec:Orientational_Dynamics}

We now examine this problem using a scaling analysis of the competition between the field- and diffusion-driven rotation for a rod fixed in space at CM position $r$. The rod's free rotational relaxation time is roughly the time it needs to diffuse over half its own length\cite{doiRotationalRelaxationTime1975}, and thus scales like $\tau_\theta \sim L^2/D$. When $r \gg L$, the force driving rotation is $F_e \sim \mathrm{d}\psi_\theta/L\mathrm{d}\theta\sim L\lambda_e kT/r^3$, and the corresponding time scale is $\tau_{e} \sim L/(F_e/\xi)$, where $\xi = k_BT/D$ is the friction coefficient. When $\tau_\theta < \tau_e$, rotational diffusion dominates and the electric forces are not sufficient to align the rod along the local field line; when $\tau_\theta>\tau_e$, on the other hand, rotational diffusion cannot stop the rod from orienting. The location $r$ where $\tau_\theta=\tau_e(r)$ scales like $r \sim {(\lambda_eL^2)^{1/3}} \sim R_\theta$, in agreement with the analysis of the equilibrium limit presented in the previous section.

\section{SIMULATIONS: METHODS AND RESULTS}
\label{sec:results}

\subsection{Coarse-grained stiff rod-like molecules}
We employ Langevin Dynamics (LD) simulations, and more precisely ESPResSo's standard coarse-grained bead-spring model\cite{slaterModelingSeparationMacromolecules2009,weikESPResSoExtensibleSoftware2019}. The excluded volume interactions between monomer beads, and between the wall and the monomers, are modeled using a repulsive Weeks-chandler-Andersen potential (WCA)\cite{weeksRoleRepulsiveForces1971}
\begin{equation}
U_\mathrm{WCA} (r) = \label{EQ:WCA}
\begin{cases}
4 \epsilon \left[ \left( \frac{\sigma}{r}\right)^{12} - \left( \frac{\sigma}{r} \right)^6 \right] +\epsilon   &\text{for } r < r_c  \\
0 &\text{for } r \geq r_c.
\end{cases}
\end{equation}
The parameter $\epsilon=k_BT$ is used as the fundamental unit of energy in our simulations, the nominal monomer size $\sigma$ is used as the fundamental unit of length, and $r_c=2^{1/6}\,\sigma$ is the cutoff length that makes $U_{WCA}$ purely repulsive.
Adjacent monomers are connected with the Finitely-Extensible-Nonlinear-Elastic (FENE) potential\cite{grestMolecularDynamicsSimulation1986}
\begin{equation}
U_{\mathrm{FENE}}(r) = - \tfrac{1}{2} K_{\mathrm{FENE}}~ r^2_0~ \textrm{ln} \left( 1 - \frac{r^2}{r_0^2} \right).
\end{equation}
We use the spring constant $K_{\textrm{FENE}}=30\, \epsilon / \sigma^2$ and the maximum extension $r_0=1.5\,\sigma$.
We control the chain stiffness via the harmonic angular potential 
\begin{equation}
U_{\textrm{Bend}}(\phi) = \tfrac{1}{2} K_{\textrm{Bend}} \left(\phi - \pi\right)^2;    
\end{equation}
with the bending constant $K_{\textrm{Bend}}=100\,\epsilon$, the molecule's persistence length is approximately equal to the nominal thermal bending length in free solution\cite{seanLangevinDynamcisSimulations2017}: $L_p/\sigma\approx{K_{\textrm{Bend}}}/{k_BT}=100$, where $\sigma=\frac{2}{3}\,r_p$ is nominal monomer size. Our five bead molecule has a contour length of $L=5\,\sigma=\frac{10}{3}\,r_p \ll L_p$.
\subsection{Langevin Dynamics Simulations}
Since the solvent is implicit in LD formalism, the equation of motion for a monomer of mass $m$ is\cite{slaterModelingSeparationMacromolecules2009}
\begin{equation}
m \dot{\vec{v}} = \vec{\nabla} U (\vec{r}) - \xi \vec{v} + \sqrt{ 2 \xi k_\mathrm{B} T } ~\vec{R}(t),
\label{Eq_lD}
\end{equation}
where $U(\vec{r}) \! = \!  U_\mathrm{WCA} (\vec{r})+U_\mathrm{FENE} (\vec{r})+U_\mathrm{Bend} (\vec{r})+U_\mathrm{E} (\vec{r})$ is the sum of the conservative potentials, with $U_\mathrm{E} (\vec{r})=QV(\vec{r})$, and $-\xi \vec{v}$ is the damping force. The last term on the \textit{rhs} is the uncorrelated noise that models the random kicks from the solvent; as usual, $\vec{R}(t)$ satisfies $\langle R_i(t) \rangle = 0$ and
$\langle R_i(t) R_j(0) \rangle = \delta(t) \delta_{ij}$, 
where $\delta (t)$ is the Dirac delta function  and $i$, $j$ represent the Cartesian coordinates.

\subsection{Static orientation}
\begin{figure}[ht]
\centering
\includegraphics{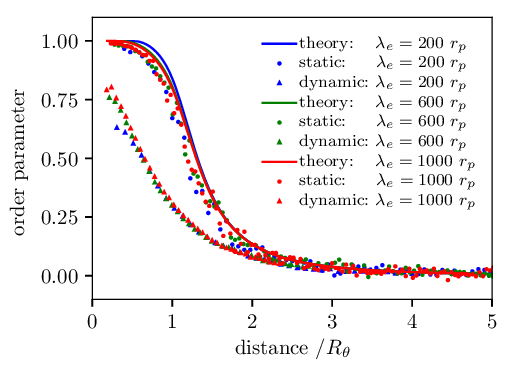}
\caption{Order parameter $\Theta(r)$ vs scaled distance $r/R_\theta$ for rods of length $L=\tfrac{10}{3}r_p$ and three field intensities $\lambda_e$ (in units of $r_p$). The data points are from simulations while the solid lines (theory) are from eqs. \ref{orderpara} and \ref{eq_avarageCos}. We show results for fixed rods that are in equilibrium (marked theory and static) as well as for free rods moving towards the pore (dynamic).}

\label{fig:simulation_rotation}
\end{figure}

We first test the theory for static orientation in Section \ref{sec:static_theory} by simulating the equilibrium orientation of the rod-like molecules when their CM position is placed at different distances $r$ right above the pore. The numerical results are in good agreement with theory, as shown in Fig.~\ref{fig:simulation_rotation}; the small deviations found for small values of $r/R_\theta$ are due to the non-radial field lines near the nanopore as discussed in our previous paper\cite{qiaoVoltageDrivenTranslocationDefining2019}. 

\begin{figure}[ht]
\centering
\includegraphics{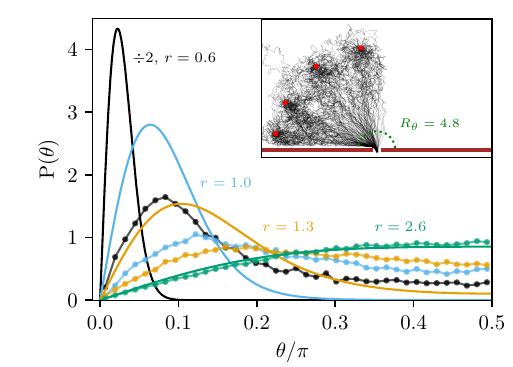}
\caption{Normalized probability distribution functions $P(\theta)$ for the local rod orientation $\theta$ at different radial distances from the pore (r=0.6, 1, 1.3 and 2.6 $R_\theta$); the field intensity is $\lambda_e=600\,r_p$, the orientational capture radius is $R_\theta=4.8\, r_p$ and the rods are launched from an initial radial distance of 27.6 $r_p$. The connected data points are from LD simulations (ensemble size of 40,000); each point corresponds to a bin size of $\pi/80$. The solid lines give the equilibrium distribution function $P(\theta) \sim \sin \theta \exp{(-\delta \Psi_\theta(r)/k_BT)}$. Inset: Sample trajectories for rods launched from four different initial angles (red dots; the angles, from the wall, are $=\frac{\pi}{2}j$; $j=0.1,\,0.3,\,0.6$ and $0.9$). The green dashed line shows the orientational radius $R_\theta$.}
\label{fig:an_distribution}
\end{figure}

\subsection{Orientation during capture}

We now simulate the capture of a randomly oriented rod released far from the pore ($\lambda_e > r \gg R_\theta$) and evaluate its mean orientation $\Theta(r)$ from an ensemble of 5000 trajectories. How a rod-like molecule enters a pore depends on multiple factors, such as the pore-molecule interactions and the detailed field lines, but since this is not our focus here, we stop the simulation once the rod is at a distance $r=L/2$ away from the nanopore. 

As shown in Fig.~\ref{fig:simulation_rotation}, the dynamic order parameter curves $\Theta(r)$ also collapse if $r$ is rescaled by $R_\theta(\lambda_e)$, implying that $R_\theta$ is again the relevant length scale. For distances $r<3\,R_\theta$, however, the drift towards the pore is too fast for the rod to adapt to the local field conditions and the order parameter is less than predicted by equilibrium theory for all three field intensities. The deviations can be also observed when looking at the orientation probability distribution function:   Fig.~\ref{fig:an_distribution} shows that the rods have a much larger probability of being unaligned than what is predicted by the equilibrium theory. The inset of Fig.~\ref{fig:an_distribution} shows some capture trajectories with rods starting from different polar angles; because of the radial symmetry of the field (except near the pore), the trajectories and orientation statistics of the rods do not depend on this angle (note however that the rods starting very close to the wall are affected by the steric restrictions to rotational motion).

The fact that $R_\theta$ is also the relevant length scale for dynamic orientation can be  understood as follows for a LD model. The time needed for the CM of the rod to move over a distance $\Delta r$ is simply $\Delta t=\Delta r/v(r)$. The amount of rotation achieved during that time is $\Delta \theta \sim \Delta t/\tau_e$; using the expressions for $\Delta t$ and $\tau_e$ given previously, we obtain the simple scaling $\Delta \theta / \Delta r\sim  1/r$. This has to be compared to the expected difference in equilibrium orientation, $\partial \langle \theta \rangle / \partial r$. The latter can be calculated in the $r \gg L$ limit using the approach given in eqs. \ref{eq_rotation_V_Gary2} - \ref{eq_avarageCos}, giving $\partial \langle \theta \rangle / \partial r \sim \lambda_e L^2/r^4$. These two rates are equal at a distance $r \sim(\lambda_eL^2)^{1/3} \sim R_\theta $. In other words, rod orientation is never in equilibrium during capture.

\subsection{Effect of initial orientation}
\label{Sec:initial_orien}

\begin{figure}[ht]
\centering
\includegraphics{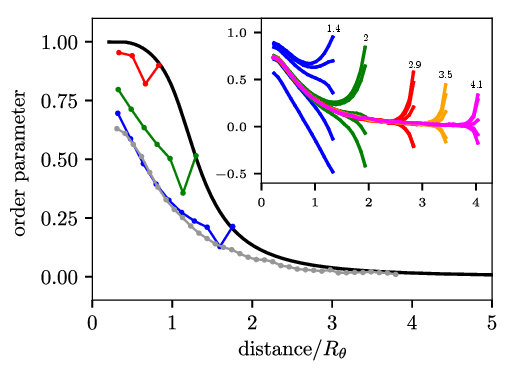}
\caption{Order parameter $\Theta(r)$ vs scaled distance $r/R_\theta$ for a field intensity $\lambda_e=1000\,r_p$. Main figure: Rods are launched from four different vertical initial positions $r$  after their orientation has reached local equilibrium (given by the black curve). Inset: The rods are launched from five different initial positions $r$ right above the nanopore and five different orientations ($\tfrac{i}{8}\pi;~i=0,1,2,3,4$). All curves are averaged over 5000 trajectories.}
\label{Fig:inital_eq}
\end{figure}

Figure~\ref{Fig:inital_eq} shows that the capture process is impacted by the fact that rod orientation is not in equilibrium with the local field if $r \lesssim R_\theta$. {Molecules are first released from different initial vertical positions $r$ with initial orientations in local static equilibrium (note that starting the rod along another polar angle gives the same results because of the radial symmetry of the field lines -- data not shown).} The black curve is the static order parameter (same as inset of Fig.~\ref{Fig_static_rotation}), while the gray data points show a case with the initial position $r \gg R_\theta$. Although the three curves corresponding to smaller initial distances tend to converge towards the gray curve, only one does so before reaching the pore. In the other two cases, \textit{i.e.} for $r \lesssim 1.5 R_\theta$, the initial rod orientations impact the entire capture trajectory. 

The memory effects are more obvious if we launch the rods with specific initial angles $\theta$. For the inset of Fig~\ref{Fig:inital_eq}, we chose five different initial orientations, from perfectly aligned with, to orthogonal to, the local field lines. When the rods start at distances $r \gtrsim 1.5 R_\theta$, the initial orientation is rapidly lost and all trajectories converge to the one shown in grey in the main figure: reorientation is thus faster than capture. However, when the rods start at distances $r \lesssim 1.5 R_\theta$, the initial orientation affects the capture process and the curves do not merge before capture: the capture time of the rods then depend on both their initial position and orientation.

\section{Conclusions}
Using scaling arguments, equilibrium calculations and LD simulations, we showed that we need a second length scale (besides the nominal capture radius $\lambda_e$), the \textit{orientational capture radius} $R_\theta=\sqrt[3]{\lambda_eL^2/60}$, in order to describe the capture of charged rods of length $L$ by a nanopore. First, $R_\theta$ is the radial distance at which rods start orienting if their initial radial position is $r_o>R_\theta$. However, because rotational dynamics is slower than capture when $r_o<R_\theta$, the rods orient less than predicted by local equilibrium arguments. The last point also implies that if $r_o<R_\theta$, the final orientation of the rod at the pore does depend on its initial orientation. While the main part of Fig.~\ref{Fig:inital_eq} shows that rods starting at distances $r_o<R_\theta$ are on average more oriented when they reach the pore than those starting further, the inset shows that some actually orient less. These results must be taken into account when studying how rods enter nanopores.

The new length scale $R_\theta$ includes the two relevant lengths in this problem, the rod length $L$ and the nominal capture radius $\lambda_e$. Under normal experimental conditions, one would have $\lambda_e \gg R_\theta \gg r_p$ and the rods are captured well before they orient. However, unless one uses high field intensities $\lambda_e \gg 60 $, the value of $R_\theta$ may not be much larger than the length of the rod itself. 

As shown above, the capture of a rod is affected by its initial orientation if $r_o < R_\theta$. To estimate the maximum impact on the capture time, let's consider two non-rotating rods, one starting parallel ($\parallel$) to the local field lines and the other starting perpendicular ($\perp$). The mean capture time of a such a rod starting from distance $r< R_\theta$ would be $\tau_E(r)=\int_0^r dz/v(z)= r^3/\lambda_e D$. Since $D_\parallel \approx 2 D_\perp$ for a rod, the difference in arrival times would be at most a factor of 2. However, since $\tau_E(R_\theta)/\tau_E(\lambda_e)=(R_\theta/\lambda_e)^3 \ll 1$, this is not expected to be important during experiments, unless one can manipulate rod orientations prior to, or during the experiment.

Our theoretical analysis and LD simulations neglect all hydrodynamics/electrohydrodynamics effects; the latter are necessary to properly model the electrophoresis of a charged rod (e.g., the effective charge $Q$ depends on ion concentration and the rod's aspect ratio\cite{grosbergDNACaptureNanopore2010,allisonModelingFreeSolution1998,allisonDependenceElectrophoreticMobility2010,allisonLengthDependenceTranslational2001}). More importantly, the friction coefficient of the rod is independent of its orientation in LD, with direct impact on rotation and capture times. In the case of flexible polymers, the electric field will orient \textit{and} deform the molecules, with impact on the capture and translocation processes\cite{farahpourChainDeformationTranslocation2013,vollmerTranslocationNonequilibriumProcess2016}. For nonlinear polymers and/or non-uniform charge distributions, the electric forces might orient the object very differently. These subtle issues will be addressed in future papers.

\begin{acknowledgements}
Simulations were performed using ESPResSo 4.0\cite{weikESPResSoExtensibleSoftware2019} on Compute Canada's Cedar system. GWS acknowledges the support of both the University of Ottawa and the Natural Sciences and Engineering Research Council of Canada (NSERC), funding reference number RGPIN/046434-2013. LQ is supported by the Chinese Scholarship Council and the University of Ottawa. 
\end{acknowledgements}
\nocite{*}
\bibliography{Refs}
\end{document}